\documentclass{aastex}
\usepackage{emulateapj5}
\usepackage{epsfig}

\newcommand{\target}{XTE\,J1118+480}

\slugcomment{Submitted to Astrophysical Journal Letters}

\shorttitle{Multiwavelength observations of XTE J1118+480}
\shortauthors{Hynes et al.}

\begin{document}

\title{The X-ray transient XTE J1118+480: Multiwavelength
observations of a low-state mini-outburst}

\author{R. I. Hynes\altaffilmark{1}, C. W. Mauche\altaffilmark{2},
C. A. Haswell\altaffilmark{3}, C. R. Shrader\altaffilmark{4},
W. Cui\altaffilmark{5,6}, and S. Chaty\altaffilmark{3}}

\altaffiltext{1}{Department of Physics and Astronomy, University of
Southampton, Southampton, SO17 1BJ, UK; rih@astro.soton.ac.uk.}
\altaffiltext{2}{Lawrence Livermore National Laboratory, L-43, 
7000 East Avenue, Livermore, CA 94550, USA; mauche@cygnus.llnl.gov.}
\altaffiltext{3}{Department of Physics and Astronomy, The Open
University, Walton Hall, Milton Keynes, MK7 6AA, UK;
c.a.haswell@open.ac.uk, s.chaty@open.ac.uk.}
\altaffiltext{4}{Laboratory for High-Energy Astrophysics, NASA
Goddard Space Flight Center, Greenbelt, MD 20771,
USA; shrader@grossc.gsfc.nasa.gov.}
\altaffiltext{5}{Center for Space Research, Massachusetts Institute of
Technology, Cambridge, MA 02139, USA; cui@space.mit.edu.}
\altaffiltext{6}{Present address: 
Department of Physics, Purdue University, West
Lafayette, IN 47907, USA}

\begin{abstract}
We present multiwavelength observations of the newly discovered X-ray
transient \target\ obtained in the rising phase of the 2000 April
outburst.  This source is located at unusually high Galactic latitude
and in a very low absorption line of sight.  This made the first {\it
EUVE} spectroscopy of an X-ray transient outburst possible.  Together
with our {\it HST}, {\it RXTE}, and UKIRT data this gives
unprecedented spectral coverage.  We find the source in the low hard
state.  The flat IR--UV continuum appears to be a combination of
optically thick disk emission and possibly synchrotron, while at
higher energies, including EUV, a typical low hard state power-law is
seen.  {\it EUVE} observations reveal no periodic modulation,
suggesting an inclination low enough that no obscuration by the disk
rim occurs.  We discuss the nature of the source and this outburst and
conclude that it may be more akin to mini-outbursts seen in
GRO\,J0422+32 than to a normal X-ray transient outburst.
\end{abstract}

%
%%%%%%%%%%%%%%%%%%%%%%%%%%%%%%%%%%%%%%%%%%%%%%%%%%%%%%%%%%%%%%%%%%%%%%%%%%%%%%%
%
\section{Introduction}
\label{IntroSection}
Soft X-ray transients (SXTs), also known as X-ray novae, (Tanaka \&
Shibazaki 1996) are low-mass X-ray binaries in which long periods of
quiescence, typically decades, are punctuated by very dramatic X-ray
and optical outbursts, often accompanied by radio activity.  In a
prototypical outburst, the luminosity approaches the Eddington limit
and X-ray emission is dominated by thermal emission from the hot inner
accretion disk.  Optical emission is then thought to be produced by
reprocessing of X-rays.  There are exceptions, however, and some
outbursts never show disk X-ray emission (e.g., GRO\,J0422+32; see
Nowak 1995 for summary and discussion).

\target\ was discovered by {\it RXTE} on 2000 March 29 (Remillard et
al.\ 2000) as a weak, slowly rising X-ray source.  Analysis of earlier
data revealed an outburst in 2000 January reaching a similar
brightness.  A power-law spectrum was seen to at least 120\,keV
(Wilson \& McCollough 2000), with spectral index similar to Cyg X-1 in
the low hard state.  A 13th magnitude optical counterpart was promptly
identified, coincident with an object with red magnitude 18.8 in Sky
Survey images (Uemura, Kato, \& Yamaoka 2000; Uemura et al.\ 2000).
The optical spectrum was typical of X-ray novae in outburst (Garcia et
al.\ 2000).  Continued observations revealed a weak photometric
modulation on a 4.1\,hr period (Cook et al.\ 2000).  The optical
brightness is surprising, as the X-rays are so faint.  It was
suggested that the system might be at very high inclination, so that
the X-ray source was obscured by the disk rim and only scattered
X-rays are visible (Garcia et al.\ 2000).  A radio counterpart was
also discovered with flux 6.2\,mJy (Pooley \& Waldram 2000).

\target\ has a very high Galactic latitude ($b=+62\degr$) and is close
to the Lockman hole (Lockman, Jahoda, \& McCammon 1986).  Consequently
it has a very low interstellar absorption of $E(B-V) \la 0.024$
(Garcia et al.\ 2000).  This, together with its brightness, make
\target\ an ideal target for multiwavelength coverage.  The
observations described here were performed as part of a coordinated
multi-wavelength campaign.  A highlight of the campaign is that the
low interstellar absorption allowed the first {\it EUVE} spectroscopy
of an X-ray transient.  This campaign is still underway, so we report
here only our earliest observations of 2000 April 4--18.
%
%%%%%%%%%%%%%%%%%%%%%%%%%%%%%%%%%%%%%%%%%%%%%%%%%%%%%%%%%%%%%%%%%%%%%%%%%%%%%%%
%
\section{{\it EUVE} data}
\label{LCSect}
{\it Extreme Ultraviolet Explorer\/} ({\it EUVE\/}) observations of
\target\ took place during 2000 April 8.10--8.71, 13.32--13.93,
and 16.91--19.60 UT.  For a description of the {\it EUVE\/} satellite
and instrumentation, refer to Bowyer \& Malina (1991), Abbott et al.\
(1996), and Sirk et al.\ (1997).  The deep survey (DS) photometer
achieved net exposures of 0.93, 0.90, and 6.92\,ks, limited by
shutdowns due to high background levels.  The short wavelength (SW)
spectrometer was not affected by such shutdowns and achieved net
exposures of 19.4, 19.8, and 81.1\,ks.  The bandpasses of the DS
photometer and SW spectrometer are defined by a Lexan/Boron filter and
extend from $\approx 70$--180\,\AA\ although interstellar absorption
extinguishes the flux longward of about 120\,\AA.  The 2nd and 3rd
observations were offset-pointed to recover more of the
short-wavelength flux.

The mean background-subtracted DS count rates for the three visits do
not vary by much: 1.61, 1.51, and 1.62\,cnts\,s$^{-1}$.  Individual
light curves show statistically significant variability about the mean
(e.g., Fig.\ 1; $\chi^2/{\rm dof}=99.3/41=2.42$).  To determine if
this variability is correlated with orbital phase, we folded the data
from the April 18 observation on the optical photometric period of
0.17082\,d (Patterson 2000), with $T_0=647.663\pm0.004$ (HJD)
corresponding to optical maximum (J. Patterson, private
communication).  The resultant light curve is shown in the inset of
Fig.\ 1.  There is no evidence for a modulation in the EUV flux of
greater than 6\,\%\ full amplitude on this period.  We also performed
a period search on these data but found no period significant at the
90\,\%\ level, for $3\,{\rm hr} < P < 4\,{\rm day}$.

\noindent
\epsfig{angle=90,width=82mm,file=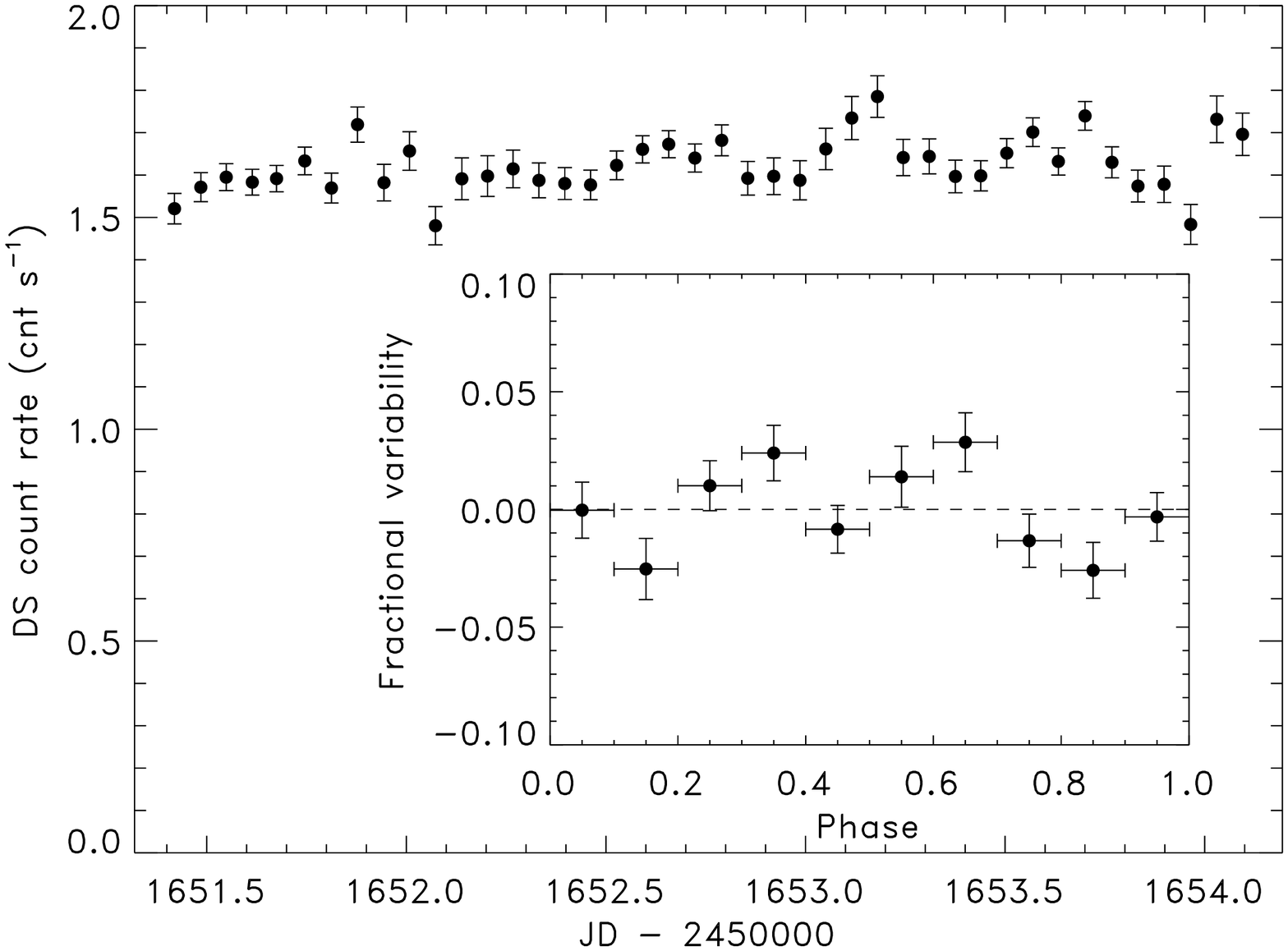}
%\figcaption{

{\footnotesize Fig.\ 1.---{\it EUVE}/DS light curve from 2000 April
18.  The error bars are the $1\, \sigma $ count rate errors from the
photon statistics.  The inset shows the same data folded on the period
of Patterson (2000), binned into 10 phase bins and converted to
fractional variations about the mean.}
\label{EUVELCFig}
%}
\vspace{1mm}

Figure 2 shows the background-subtracted mean {\it EUVE} SW spectrum
from the April 18 observation binned to $\Delta\lambda = 0.5$\,\AA,
matching the spectral resolution of the instrument, and removing the
nonstatistical correlation between neighboring wavelength bins. The
short-wavelength limit of the spectrum is dictated by the strong
increase in the background shortward of $\sim 70$~\AA , while
interstellar photoelectric absorption suppresses the flux longward of
$\sim 120$~\AA.  To approximately constrain the absorbing column
density and flux, we fit these data with constant $\nu F_{\nu}$ and
photoelectric absorption. For the latter, we used the EUV absorption
cross sections of Rumph, Bowyer, \& Vennes (1994) for \ion{H}{1},
\ion{He}{1}, and \ion{He}{2} with abundance ratios 1:0.1:0.01, typical
of the diffuse interstellar medium.
%The resulting fit parameters and 90\% confidence
%errors are $F_\lambda = (4.0^{+0.5}_{-0.3})\times 10^{-12}~\rm
%erg~cm^{-2}~s^{-1}~\AA ^{-1}$ and $N_{\rm H}= (8.7\pm0.4)\times
%10^{19}~\rm cm^{-2}$, with $\rm \chi^2/dof=185.5/105=1.77$.  
The resulting fit parameters and 90\% confidence errors are $\nu F_\nu
= (2.5^{+0.2}_{-0.3})\times 10^{-10}~\rm erg~cm^{-2}~s^{-1}$ and
$N_{\rm H}=(7.4\pm 0.4)\times 10^{19}~\rm cm^{-2}$, with $\rm
\chi^2/dof=206.2/105=1.96$.  The fit is poor because the data
systematically fall below this simple model shortward of $\sim
75$\,\AA, either because of a flattening of the continuum spectrum, or
an absorption feature or edge.  There may also be a broad emission
feature at $\sim81$\,\AA.  We cannot usefully constrain the slope of
the EUV spectrum but assuming a wide range of intrinsic EUV spectra,
$F_\nu \propto \nu^{\alpha}$, where $+2 \geq \alpha \geq -4$, gives a
range of column densities of (0.35--1.15)\,$\times10^{20}$\,cm$^{-2}$
respectively.  These values are relatively low, and are fully
consistent with the expected column density of the interstellar medium
in the vicinity of the Lockman hole,
(0.5--1.5)\,$\times10^{20}$\,cm$^{-2}$ (B. Y. Welsh, private
communication).  There does not therefore appear to be any strong
absorption intrinsic to the source, although estimates from fits to
the {\it EUVE} spectrum are sensitive to the assumed ionization state
of the absorber.

\noindent
\epsfig{angle=90,width=82mm,file=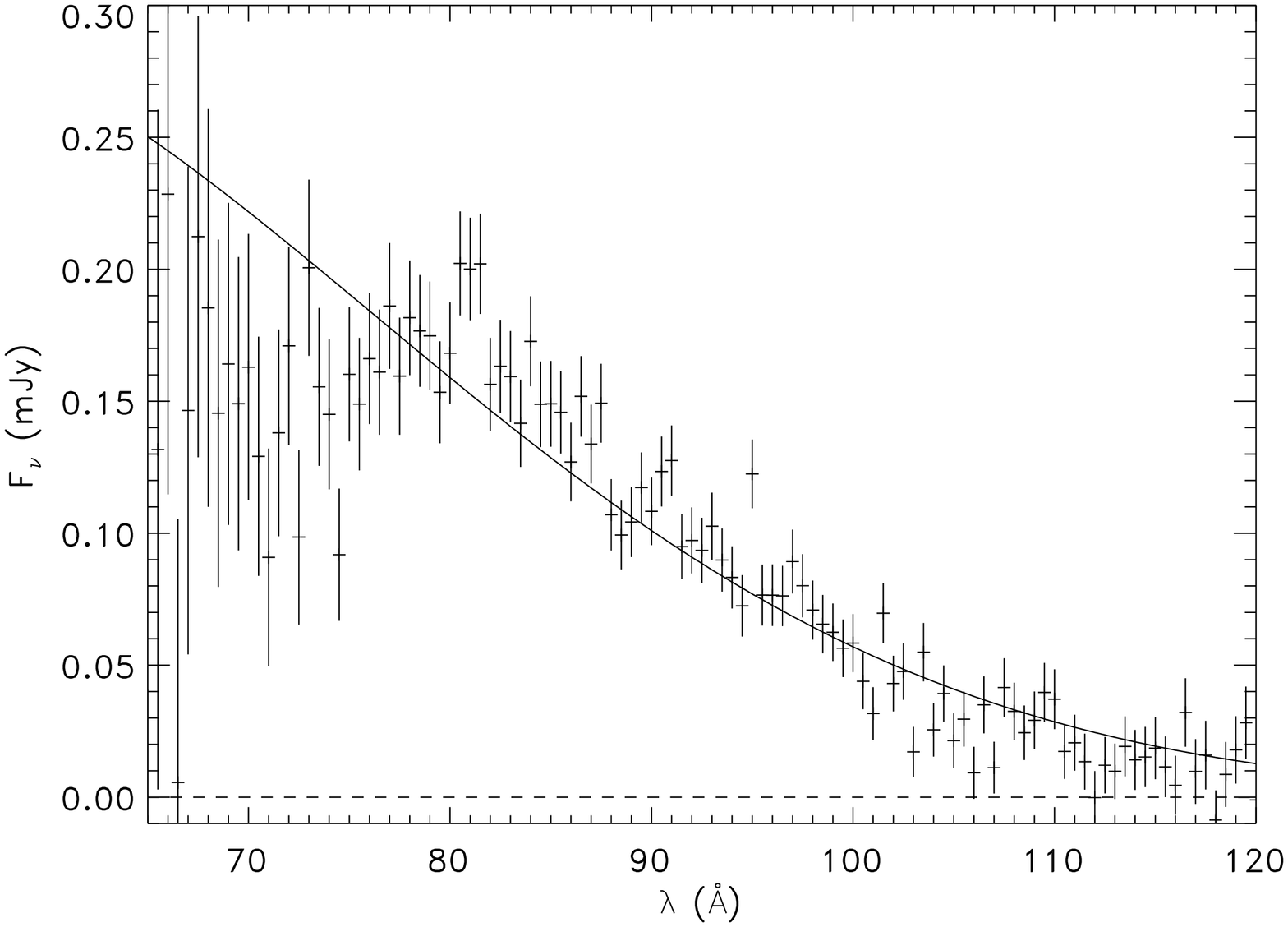}
%\figcaption{

{\footnotesize Fig.\ 2.---Mean {\it EUVE}/SW spectrum from 2000 April
18.  The vertical error bars are the $1\, \sigma $ errors from the
photon statistics.  The fit is constant $\nu F_{\nu}$ subject to
photoelectric absorption.}
\label{EUVESpecFig}
\vspace{1mm}
%}

The low intrinsic absorption implied by the EUV spectrum together with
the lack of orbital modulation is evidence against a high inclination
explanation for the low X-ray to optical flux ratio (Garcia et al.\
2000).  In that case, because of the extreme sensitivity of the EUV
flux to absorbing material, we would expect some modulation due to
asymmetry in the disk rim and significant intrinsic absorption by
material above the disk.  It instead appears likely that the central
disk regions are unobscured, and that the system is in an intrinsically
X-ray faint state.  This is also consistent with the lack of optical
eclipses.  This argument does, however, assume that EUV emission
originates from the disk.  If it actually comes from a wind, as in the
cataclysmic variable OY Car (Mauche \& Raymond 2000), for example,
then the system could still be at high inclination.  
%
%%%%%%%%%%%%%%%%%%%%%%%%%%%%%%%%%%%%%%%%%%%%%%%%%%%%%%%%%%%%%%%%%%%%%%%%%%%%%%
%
\section{{\it HST} and UKIRT data}
{\it Hubble Space Telescope (HST)} observations were performed with
the Space Telescope Imaging Spectrograph (STIS) on 2000 April
8.52--8.81 and 18.57--18.74 UT using the E140M, E230M, G430L, and
G750L modes.  An average calibrated spectrum for April 8 was
constructed from standard {\it HST} pipeline data products, with
useful coverage from 1150--10000\,\AA.  The region from
1195--1260\,\AA\ was excluded as this was completely dominated by
\ion{N}{5}\ emission and Ly$\alpha$ absorption (Haswell, Hynes, \&
King 2000a).

Near-infrared observations were carried out at the UKIRT 3.8\,m
telescope using UFTI (1--2.5\,$\mu$m) and IRCAM/TUFTI (1--5\,$\mu$m)
on 2000 April 4 and 18.  Images were obtained in $JHK$ with UFTI and
$JHKL'M'$ with IRCAM/TUFTI.  Exposure times were 10--60\,s and the
conditions were photometric.  The images were processed by removal of
the dark current, flat-fielding, and sky subtraction.  The magnitudes
acquired on April 4.2 UT with UFTI are: $J = 12.12\pm0.02$, $H =
11.75\pm0.02$, and $K = 11.06\pm0.02$ (Chaty et al.\ 2000).  The
magnitudes acquired on April 18.5 UT with IRCAM/TUFTI are: $J =
11.92\pm0.07$, $H = 11.43\pm0.06$, $K = 11.05\pm0.08$, $L' =
9.71\pm0.14$, and $M' = 9.38\pm0.42$.  The source brightened between
these two dates, which bracket the dates of {\it HST} observations
described above.

From UV to IR wavelengths ($\sim1000$--50000\,\AA) the spectrum is
flat with $F_{\nu} \approx 24$\,mJy, constant to within 10\,\%,
although a weak Balmer jump in absorption of $\sim7$\,\% is present.
The latter likely indicates that some optically thick disk emission is
present.  The underlying flat continuum, however, does not resemble a
disk spectrum, as is discussed in Section 5.
%
%%%%%%%%%%%%%%%%%%%%%%%%%%%%%%%%%%%%%%%%%%%%%%%%%%%%%%%%%%%%%%%%%%%%%%%%%%%%%%
%
\section{{\it RXTE} data}

We observed \target\ with the {\it Rossi X-ray Timing Explorer (RXTE)}
Proportional Counter Array (PCA) and High-Energy Timing Experiment
(HEXTE) at several epochs selected to coincide with the {\it HST}
visits. This included 2000 April 8.55--8.58 when a total exposure of
about 4\,ks was obtained. Subsequent discussion refers to that
observation.  

For the PCA, we used the ``standard mode'' data (128 spectral
channels, 16\,s accumulations), selecting sub-intervals when the
number of detectors on remained constant (about 75\,\% of the total).
The current PCA background model and/or the instrument responses are
unreliable above about 25\,keV, so we did not use those channels for
our model fitting.  There was sufficient PCA--HEXTE overlap so that this
was not a serious limitation.  The lowest several energy channels are
also known to be problematic so were discarded.  The total flux on the
2--10\,keV band was about
$1.0\times10^{-9}$\,ergs\,cm$^{-2}$\,s$^{-1}$, i.e.\ about 40\,mCrab.
For HEXTE, the exposure time was rather marginal for detailed spectral
modeling given the source intensity relative to the background, but
the source seems to be solidly detected to $\sim80$\,keV, and perhaps
above 100\,keV.

The data sets were then simultaneously deconvolved to derive a best
estimate of the incident photon flux.  Given the known
cross-calibration discrepancies between the two instruments, we
allowed the two normalization terms to vary independently.  The
spectrum is hard, with a photon power-law index of about $1.8\pm0.1$.
A thermal Comptonization model (Sunyaev \& Titarchuk 1980) with
$\tau=3.2$ and $T_{\rm e}=33$\,keV, also provided an acceptable fit.
In both cases, there was a distinct positive residual corresponding to
the 6.4\,keV Fe K resonance, thus a Gaussian line profile was included
to refine our overall fit.  Given the low hydrogen column density, and
our lack of low-energy coverage, the effect of interstellar absorption
on the X-ray spectrum is negligible.  There was no evidence (in terms
of statistical improvement to our fits) for a soft-excess component.
%
%%%%%%%%%%%%%%%%%%%%%%%%%%%%%%%%%%%%%%%%%%%%%%%%%%%%%%%%%%%%%%%%%%%%%%%%%%%%%%%
%
\section{The spectral energy distribution}
\label{SEDSect}
Figure 3 shows the spectral energy distribution (SED) inferred from
radio to hard X-ray data.  Radio data obtained with the VLA and Ryle
telescope between 2000 April 2.74--3.13 were reported by Dhawan et al.\
(2000).  Between then and April 18, the radio was no more than 30\,\%\
brighter, i.e.\ $\Delta\log F_{\nu} \la 0.1$ (G. Pooley, private
communication).  UKIRT data were obtained on April 4 and 18 and are
consistent to within $\Delta\log F_{\nu} \sim 0.15$.  The other data
were obtained simultaneously during April 8.  Calibration errors in
the {\it HST} and {\it RXTE} data are estimated at $\Delta\log F_{\nu}
\la 0.1$.  The {\it EUVE} uncertainty is dominated by the uncertain
absorption as shown.

These data have been corrected for interstellar absorption where
necessary.  Assuming the column density inferred above to be
interstellar, and adopting an average gas-dust ratio (Bohlin, Savage,
\& Drake 1978) and extinction curve (Seaton 1979), we corrected the IR--UV
data for reddening with $E(B-V)=0.013$.  This makes only a small\
difference at these wavelengths as the value is so low.  The assumed
absorption is important at EUV wavelengths so for the {\it EUVE} data we
plot three solutions with $N_{\rm H}=0.35$, 0.75, and
$1.15\times10^{20}$\,cm$^{-2}$ to illustrate how the flux and spectral
index vary with assumed absorption.

\noindent
\epsfig{angle=90,width=82mm,file=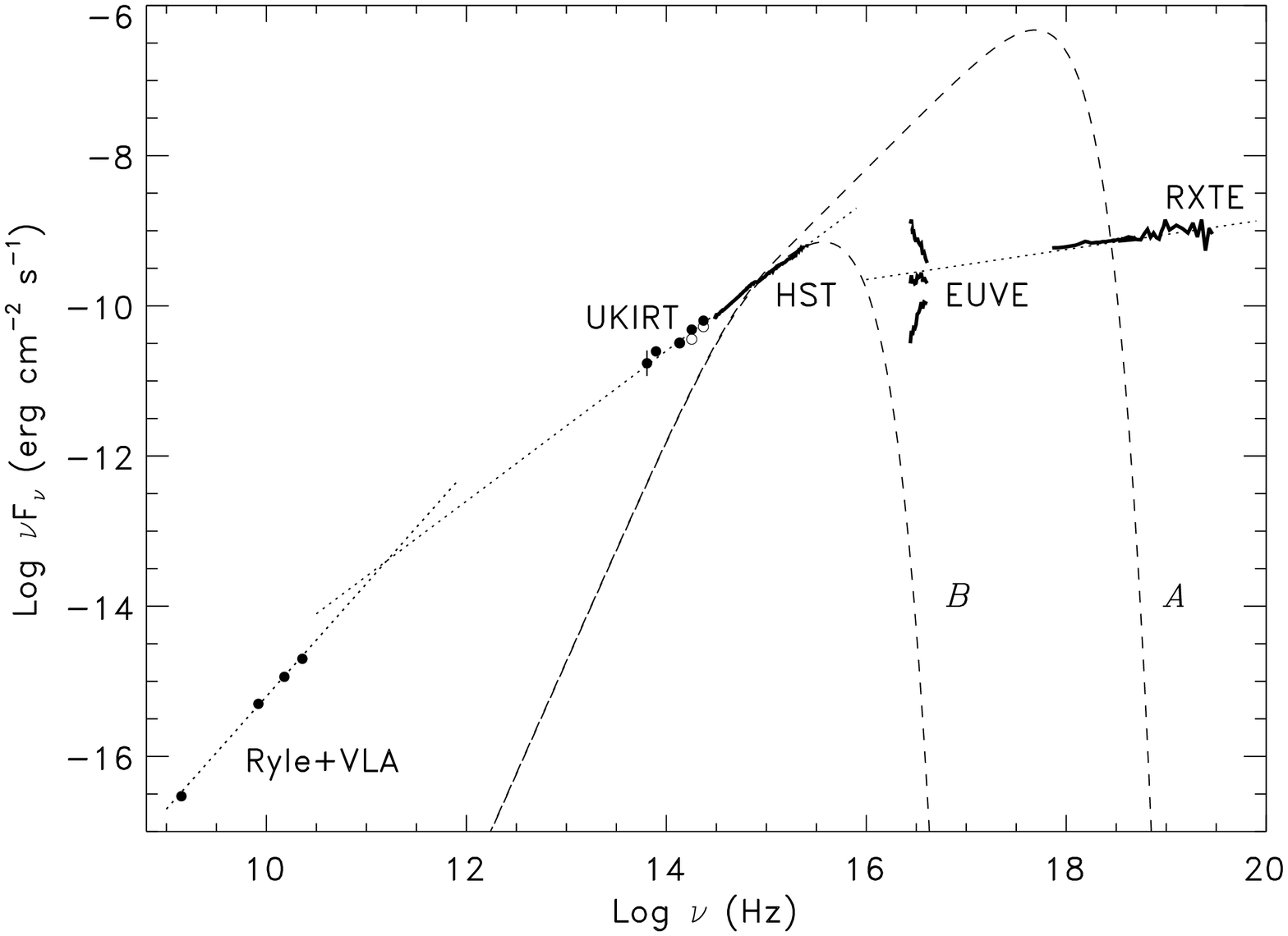}
%\figcaption{

{\footnotesize Fig.\ 3.---Spectral energy distribution from radio to
hard X-ray frequencies on 2000 April 8 ({\it HST, EUVE, RXTE}), April
2--3 (radio), and April 4, 18 (UKIRT).  {\it EUVE} data are shown
corrected for three different amounts of absorption.  Dashed lines are
two steady-state disk models with differing inner radii.  Dotted lines
are characteristic power-laws ($F_{\nu}\propto\nu^{\alpha}$; from left
to right, $\alpha=+0.5$, $0.0$, $-0.8$).  See text for details.}
\label{SEDFig}
\vspace{1mm}
%}

The presence of a Balmer jump in the {\it HST} optical data suggests
that an optically thick disk makes a significant contribution there.
Fig.\ 3 shows model SEDs calculated for a steady-state viscously
heated disk (Shakura \& Sunyaev 1973).  These are intended to be
schematic rather than detailed fits.  The low energy cut-off is fixed
by an outer temperature of 8000\,K, chosen to be consistent with the
hot phase of the disk instability model.  The high energy cut-off then
depends on the inner radius: $\sim3$\,R$_{\rm s}$ and
$\sim2000$\,R$_{\rm s}$ for models A and B respectively (these assume
an orbital period of 4.1\,hr and a 7\,M$_{\odot}$ black hole as the
accretor).  Model A is representative of a high state SED, with the
disk extending to the last stable orbit.  This is clearly ruled out by
the EUV and X-ray data as a direct consequence of the low X-ray to
optical flux ratio.  The uncertainty in the system parameters,
including the possibility that the central object is a neutron star,
cannot resolve this.  Model B is more like the low hard state scenario
proposed by Esin et al.\ (1997) in which the center of the disk (out
to 10$^3$--10$^4$\,R$_{\rm s}$) is evaporated into an advective flow.
Such a model is consistent with the low X-ray to optical flux ratio
and reproduces the UV slope fairly well.  The EUV to hard X-ray range
can be approximately fitted by a single power-law,
$F_{\nu}\propto\nu^{-0.8}$, typical of the low hard state, and so
would not be disk emission.  This spectral region does resemble the
low hard state SEDs presented by Esin et al.\ (1997), as does the
X-ray to optical flux ratio.  Alternative explanations of this low
flux ratio include material leaving the system in an outflow (e.g.,
Blandford \& Begelman 1997), or accumulating at larger radii in a
non-steady state disk; either option will reduce the central accretion
rate as required.

Disk models provide a very poor fit in the IR, as a disk spectrum
should steepen to a Rayleigh-Jeans tail well before the $M'$ band is
reached (unless the edge of the disk is at $T\la1000$\,K).  It thus
appears that another source of near-IR flux is present.  This may be
related to the radio emission, as this lies close to an extrapolation
of the IR--UV power-law.  The very flat IR--UV part of the SED
($F_{\nu} \sim {\rm constant}$) could then very naturally be
interpreted as a mixture of an optically thick disk spectrum and
flat-spectrum emission, possibly synchrotron, such as is seen at
radio--mm wavelengths in Cyg X-1 and other black hole candidates
(Fender et al.\ 2000).  The radio emission has a steeper, inverted
spectrum ($F_{\nu} \propto \nu^{0.5}$), likely optically thick
synchrotron emission.
%
%%%%%%%%%%%%%%%%%%%%%%%%%%%%%%%%%%%%%%%%%%%%%%%%%%%%%%%%%%%%%%%%%%%%%%%%%%%%%%%
%
\section{Discussion}
Consideration of the spectral energy distribution suggests that during
the period 2000 April 4--18 the source was in the low hard state.
Power density spectra at X-ray, UV, and optical wavelengths support
this, showing band-limited noise and QPOs typical of this state
(Revnivtsev, Sunyaev, \& Borozdin 2000; Haswell et al.\ 2000b).  This
in itself is not particularly striking: GRO\,J0422+32, for example,
spent its entire 1992--1993 main outburst in the low hard state.  What
is striking is the low X-ray to optical flux ratio: it is an unusually
low state.  This is not typical of a major SXT outburst.  There are,
however, similarities in this behavior to {\em mini-outbursts} in
GRO\,J0422+32 (Castro-Tirado, Ortiz, \& Gallego 1997; Shrader et al.\
1997).  Like the outbursts of \target, these outbursts were
characterized by a very low X-ray to optical flux ratio and low hard
state behavior.  The optical outbursts have comparable length and
brightness; in GRO\,J0422+32 the mini-outburst duration is $\sim60$\,d
with $\sim120$\,d recurrence time.  In \target\, the first outburst
lasted $\sim40$\,d, with the second outburst beginning $\sim60$\,d
after the start of the first.  In both cases, the peak optical
brightness is $\sim6$\,mag above quiescence, with the interoutburst
brightness about 3.5\,mag above quiescence in \target\ (Uemura et al.\
2000) and 1--3\,mag above quiescence in GRO\,J0422+32.  It thus
appears likely that the outburst of \target\ is of the same kind as
the mini-outbursts in GRO\,J0422+32, suggesting that this phenomenon
is not just an aftereffect of a main outburst, but can occur in
isolation.

The nature of the compact object remains uncertain.  The spectral
energy distribution is equivocal, as both black hole and neutron star
systems can show extended power-laws in the low hard state, with
similar photon indices (Barret et al.\ 2000 and references therein).
A distinguishing neutron star feature may be a very soft excess,
$T\sim0.5$\,keV, associated with the heated neutron star surface or
boundary layer, however, our data clearly cannot address this.  We
note that Revnivtsev et al.\ (2000) have speculated that the compact
object is a black hole based on the non-detection of high frequency
variability, which is considered a signature of a neutron star.
%
%%%%%%%%%%%%%%%%%%%%%%%%%%%%%%%%%%%%%%%%%%%%%%%%%%%%%%%%%%%%%%%%%%%%%%%%%%%%%%%
%
\acknowledgments
The {\it EUVE\/} observations were made possible by a generous grant
of Director's Discretionary Time by {\it EUVE\/} Project Manager
R.~Malina, the efforts of {\it EUVE\/} Science Planner M.~Eckert, the
staff of the {\it EUVE\/} Science Operations Center at CEA, and the
Flight Operations Team at Goddard Space Flight Center.  This work
includes observations with the NASA/ESA {\it Hubble Space Telescope},
obtained at the Space Telescope Science Institute, operated by the
Association of Universities for Research in Astronomy, Inc.\ under
NASA contract No.\ NAS5-26555.  We would like to thank the {\it HST}
and {\it RXTE} support staff for ongoing efficient support.  UKIRT is
operated by the Joint Astronomy Centre on behalf of the U.K. Particle
Physics and Astronomy Research Council.  We thank the JAC, and in
particular J. K. Davies, for their open policy and efficiency and
G. P. Smith and I. Smail for their assistance with the April 4
observation.  RIH, CAH, and SC acknowledge support from grant
F/00-180/A from the Leverhulme Trust.  CWM's contribution was
performed under the auspices of the U.S.\ Department of Energy by
University of California Lawrence Livermore National Laboratory under
contract No. W-7405-Eng-48.  WC acknowledges NASA LTSA grant
NAG5-7990.
%
%%%%%%%%%%%%%%%%%%%%%%%%%%%%%%%%%%%%%%%%%%%%%%%%%%%%%%%%%%%%%%%%%%%%%%%%%%%%%%%
%

\end{document}